# Are Aryabhata's and Galilean Relativity Equivalent?

Roopa H. Narayan

**Abstract**
This paper considers the question of whether Aryabhata's direct and indirect references to relativity of motion are equivalent to Galilean relativity. Examining verses from different sections of the text *Aryabhatiya,* we show that there is explicit mention of relativity of space and motion, although there is no explicit reference to the motion being uniform.

## 1 Introduction

Aryabhata (born 476 CE) explicitly mentions earth's rotation and its many effects such as the day-night occurrence, the long winters and summers in the polar regions [1-5]. Recently, Parakh argued [6] that Aryabhata was fully aware of relativity of motion as described in the *Aryabhatiya* (Aryabhata's book on astronomy) and that his understanding of this was similar to that of Galileo. But Parakh did not bring all the evidence to bear on this question of this similarity and he considered only 2 stanzas to make his case.

In this article, we present further arguments from *Aryabhatiya* (with two additional stanzas) for an understanding of his relativity for a proper comparison with Galileo's ideas.

For those who are not familiar with the astronomy texts of ancient India, let it be noted that these texts were written in verse and each stanza succinctly described facts, procedures or principles. For a background of these texts and the observational astronomy they were based on, see [7-9]. The fact that we will consider 4 stanzas in this article, rather than Parakh's two, implies that we will be taking in much additional evidence.

## 2 Relativity of motion

Aryabhata not only describes relativity of motion with an example of boat, which is coincidentally the very same example that Galileo uses in his work almost 1000 years later, he also implicitly mentions how the situation on earth remains unaltered as far as other processes on earth are concerned.

Parakh quotes stanzas 9 and 10 from the astronomical section of *Aryabhatiya* to describe Aryabhata's ideas.

अनुलोमगतिनौस्थः पश्यत्यचलं विलोमगं यद्वत् ।
अचलानि भानि तद्वत् समपश्चिमगानि लङ्कायाम् ॥९॥

उदयास्तमयनिमित्तं नित्यं प्रवहेण वायुना क्षिप्तः ।
लङ्कासमपश्चिमगो भपञ्जरः सग्रहो भ्रमति ॥१०॥

*Translation:* Similar to a person in a boat moving forward who sees the stationary objects on the bank of the river as moving backwards, the stationary stars at Lanka (equator) are viewed as moving westwards (9).

An illusion is created similarly that the entire structure of asterisms together with the planets is moving exactly towards the west of Lanka, being constantly driven by the provector wind, to cause their rising and setting.

*Commentary:* The first stanza speaks of the commonly felt notion that makes one feel one's experience unchanged if the motion is uniform.

By using the term "illusion," Aryabhata is emphasizing the fact that similar to an observer in a moving boat, "we"-- the observers on earth-- feel that everything outside our planet moves from east to west. In fact, it is the earth which moves circularly from west to east resulting in the sun rise and sun set. This is relative motion of all that is collectively observable from the non-stationary earth.

What is noteworthy here is that he extends relativity of motion as experience on a boat to rotational motion of the stars. In other words, whereas his reference to the motion of the boat suggests uniform motion, his reference to the motion of the stars includes rotational motion.

## 3 Relativity of space

स्वर्मेरु स्थलमध्ये नरको बडवामुखं च जलमध्ये।
अमरमरा मन्यन्ते परस्परमधःस्थितान् नियतम् ॥१२॥

*Translation:* Heavens and the Meru mountain are at the centre of the land (i.e., at the north pole); hell and the Badavamukha are at the center of the water (i.e., at the south pole). The gods (residing at the Meru mountain) and the demons (residing at the Badavamukha) consider *themselves positively and permanently below each other.*

*Commentary:* The Meru mountain is representative of the north-pole (mentioned in verse 11 of this section of the text). Aryabhata hence is stating that residents of the north and south pole consider each other as mutually being below each other. This is significant for this is an explicit declaration of relativity of space.

It is hard to estimate as to what Aryabhata's intuition was related to this relativity of space. It include recognition that there is no change in the relative position of subjects on the surface of earth and nor do they fall off. Aryabhata may or may not have known of gravitational force as an explanation for this phenomenon. But it should be noted that gravitational force is mentioned by the much earlier author Kanada in his text on atoms and is also described as a force in the early Surya Siddhanta. But in either case, this verse implies that people on the surface of the globe of earth feel in an equally privileged position. The rotation of the earth does not alter this fact.

In verse 5 & 6 of this same section of the text, earth is defined as a spherical planet suspended in space surrounded by numerous stars. It is always darker on one half which is the half that is facing away from the sun and this darkness is a consequence of its own shadow.

In verses 13 and 14 he describes how the relationship between the latitude and the time of the day varies. He further describes effects like what is clock-wise in the north pole is anti-clockwise in the south pole, and many complex calculations about eclipses.

## 4 Time and velocity

बक्रे विलोमविवरे गतियोगेनानुलोमविवरे द्वौ ।
गत्यन्तरेण लब्धौ द्वियोगकालावतीतैष्यौ ॥३१॥

*Translation:* Divide the distance between the two bodies moving in the opposite directions by the sum of their speeds, and the distance between the two bodies moving in the same direction by the difference of their speeds; the two quotients will give the time elapsed since the two bodies met or to elapse before they will meet.

*Commentary:* The velocity of a moving body on earth as a function of distance and time, is described in this verse. But this is the same as that for a moving body on a stationary earth.

Therefore it is conclusive that Aryabahta recognized that laws of motion remain the same irrespective of the motion of the reference frame.

In works like Yogavasishtha [10], and Vaisesika [11], which summarize ancient Indian ideas of physical reality, there is explicit mention of the relativity of time flow. These ideas are also found in other Indian medieval scientific texts [12].

## 5 Conclusions

Galileo (1564-1642) presented his principle as the impossibility of using "any mechanical experiment to determine absolute uniform velocity." [6]

Although there is no comparable explicit mention of this impossibility principle in Aryabhata's work, most elements that contribute to this principle are stated. In particular, there is explicit mention of relativity of space, and there is also mention of relativity as in the [uniform] motion of the boat, as well the [regular non-uniform] motion of the stars.

Aryabhata states that observers on earth do not experience their own rotational motion, observers away from the earth will detect although a westward motion. Implicitly, the laws of motion remain the same for moving objects on earth. This does sum up to the position that regular motion can be detected only by observing the system from another reference frame, a view that is virtually identical to Galilean relativity.